%% file: main.tex
\documentclass[10pt,conference]{IEEEtran}
\usepackage{cite}
\usepackage{amsmath,amssymb,amsfonts}
\usepackage{algorithmic}
\usepackage{graphicx}
\usepackage{textcomp}
\usepackage{xcolor}
\usepackage[hyphens]{url}
\usepackage{fancyhdr}
\usepackage[bookmarks=true,breaklinks=true,letterpaper=true,colorlinks,citecolor=blue,linkcolor=blue,urlcolor=blue]{hyperref}

\usepackage{cite}
\usepackage{amsmath,amssymb,amsfonts}
\usepackage{algorithmic}
\usepackage{graphicx}
\usepackage{textcomp}
\usepackage{xcolor}
\usepackage[hyphens]{url}

\usepackage{mathptmx} 
\usepackage{fancyhdr}
\usepackage[normalem]{ulem}
\usepackage[hyphens]{url}

\usepackage[bookmarks=true,breaklinks=true,letterpaper=true,colorlinks,citecolor=blue,linkcolor=blue,urlcolor=blue]{hyperref}

\usepackage{algorithmic}
\usepackage{graphicx}
\usepackage{textcomp}

\usepackage{tikz}
\usepackage{comment}
\usepackage{caption}
\usepackage{subcaption}
\usepackage{float}
\usepackage{pifont}
\usepackage{color}
\usepackage{soul}
\usepackage{enumitem}
\usepackage{mdframed}   
\usepackage{array}
\usepackage{bold-extra}
\usepackage{svg}
\usepackage{booktabs}

\definecolor{light-gray}{gray}{0.90}
\sethlcolor{light-gray}

\newcommand{\TOOL}{\texttt{iMIV}}
\newcommand{\cmark}{\ding{51}}
\newcommand{\xmark}{\ding{55}}

\newcolumntype{L}[1]{>{\raggedright\let\newline\\\arraybackslash\hspace{0pt}}m{#1}}
\newcolumntype{C}[1]{>{\centering\let\newline\\\arraybackslash\hspace{0pt}}m{#1}}
\newcolumntype{R}[1]{>{\raggedleft\let\newline\\\arraybackslash\hspace{0pt}}m{#1}}

\usepackage{ntheorem}   
\theoremstyle{nonumberbreak}
\theoremheaderfont{\bfseries}
\theorembodyfont{\normalfont}
\newmdtheoremenv[%
linecolor=black,%
linewidth=0.7pt,%
backgroundcolor=gray!40,%
innertopmargin=5pt,%
ntheorem]{formalbox}{ }

\newcommand{\ignore}[1]{}

\newcommand*\circled[1]{\tikz[baseline=(char.base)]{
            \node[shape=circle,draw,inner sep=1pt] (char) {#1};}}

\title{iMIV: \underline{i}n-\underline{M}emory \underline{I}ntegrity \underline{V}erification for NVM}

\begin{document}

\author{\normalfont
  \begin{tabular}{c @{\hskip 1.75in} c}
    \begin{tabular}{c}
    {Rajat Jain} \\
    Indian Institute of Science \\
    Bengaluru, India
    \end{tabular}
 &
    \begin{tabular}{c}
    {Aravinda Prasad}\\
    Processor Architecture Research Lab\\Intel Labs, India
    \end{tabular}
   
    \\[0.85cm]
  
    \begin{tabular}{c}
    {Sreenivas Subramoney}\\
    Processor Architecture Research Lab\\
    Intel Labs, India
    \end{tabular}
 &
    \begin{tabular}{c}
    {Arkaprava Basu}\\
       Indian Institute of Science \\
    Bengaluru, India
    \end{tabular}
    \end{tabular}
    
    }
    
\maketitle


\input{PaperContent/1_abstract}

\input{PaperContent/2_introduction}

\input{PaperContent/3_background}

\input{PaperContent/4_Security_guarantees}
\input{PaperContent/5_motivation}

\input{PaperContent/6_Key_Ideas}
\input{PaperContent/7_design}

\input{PaperContent/8_methodology}

\input{PaperContent/9_evaluation}

\input{PaperContent/10_Related_works}
\input{PaperContent/11_conclusion}


\bibliographystyle{IEEEtranS}
\bibliography{refs}

\end{document}

%% file: PaperContent/1_abstract.tex
\begin{abstract}
Non-volatile Memory (NVM) could bridge the gap between memory and storage.
However, NVMs are susceptible to data remanence attacks.
Thus, multiple security metadata must persist along with the data
to protect the confidentiality and integrity of NVM-resident data. 
Persisting Bonsai Merkel Tree (BMT) nodes, critical for data integrity, 
can add significant overheads due to need to write large amounts of metadata off-chip to the bandwidth-constrained NVMs.

We propose \textit{iMIV} for low-overhead, fine-grained integrity verification through in-memory computing. 
We argue that memory-intensive integrity verification operations (BMT updates and verification) should be employed close to the NVM to limit off-chip data movement.
We design iMIV based on typical NVDIMM designs that have an onboard logic chip with a trusted encryption engine, separate from the untrusted storage media.
iMIV reduces the performance overheads from 205\% to 55\%
when integrity verification operations are offloaded to NVM compared to
when all the security operations are employed at the memory controller.

\end{abstract}

%% file: PaperContent/2_introduction.tex
\section{introduction}
\label{sec:introduction}

NVM's byte-level persistence enables developing recoverable programs where computation can be resumed after a system crash or power failure\cite{whisper,mnemosyne,dudetm,nvheap}.
However, the data retention across power cycles renders NVM's contents susceptible to various data remanence attacks, while such attacks are not possible in DRAM.

Typically, counter-mode encryption (CME) is used for its efficiency, where a counter is incremented every time the data is written from trusted processor boundaries (SoC) to the NVM.
While encryption protects confidentiality, it cannot guarantee data integrity, i.e., the user will be unable to determine if the data was tampered with or not.
To detect tampering, a hash called the Stateful Message Authentication Code (SMAC) is generated and stored along with the ciphertext.
After retrieving the ciphertext from the untrusted NVM storage media, the hash is recalculated and checked against the SMAC. A mismatch indicates data integrity failure.

An attacker, however, can still \textit{replay} older responses by snooping on the bus. 
An integrity tree, such as the Bonsai Merkel Tree (BMT), is computed over the per-block counters used in CME to guarantee freshness of counters, and thus, detect replay attacks. 
BMT root is persisted in the trusted SoC. 

When a counter block is retrieved from the NVM storage media, BMT is computed, and the root value is compared to the value on the SoC to ensure counter freshness. 
The most recent counter value for a data block, along with SMAC, guarantees data freshness.
Persisting the counter and SMAC, along with the ciphertext and updating the BMT root, guarantees the confidentiality and integrity of data.

While it is not strictly necessary to persist intermediate BMT nodes for integrity verification, failing to do so has two major drawbacks as follows.
\textcircled{\textbf{\small{1}}} If integrity check fails due to mismatch with the BMT root, then without intermediate BMT nodes it is impossible to find which data block was tampered.
Even a single tampered byte renders the \textit{entire} content of many terabytes of NVM unreliable. 
\textcircled{\textbf{\small{2}}} Upon boot after a crash or power cycle, the entire BMT must be reconstructed for integrity verification of NVM contents which can take hours on systems 
with terabytes of aggregate NVM capacity~\cite{anubis}.

Therefore, all intermediate nodes from the leaf to the root of a BMT should be persisted for each write to enable fine-grain verifiability.
Unfortunately, most prior works on secure NVM~\cite{osiris,janus,phoenix,bmf} avoid persisting intermediate BMT nodes due to high overheads.
Also, the height of BMT increases with aggregate NVM capacity on the system. 
For instance,  processors can accommodate several terabytes of aggregate NVM capacity~\cite{optane-high-end-system}, resulting in a BMT with 10 or more levels.
Triad-NVM~\cite{triad-nvm} trade-offs performance for better verifiability and shorter recovery time by partially persisting the BMT nodes, but
it can still result in gigabytes of unverifiable region in case of a BMT root mismatch.

We aim to design a secure NVM that provides fine-grained data integrity verification and confidentiality at low-performance overheads and fast recovery times.
This necessitates computing and persisting all the intermediate BMT nodes, in addition to SMAC and counters, on every write to NVM with minimal overhead.
\ignore{A complementary objective is to align our design with commercial NVDIMM architectures such as Intel's Optane DC PMM for a more significant impact.}

Toward this, we first analyze the overheads of fine-grained data integrity verification.
Unlike prior works, we find a more realistic estimate of the overheads 
by assessing \texttt{NVM-aware} applications that \textit{only} persist  the data needed for the correct recovery. 
Consequently, only a fraction of stores persist to NVMs~\cite{whisper}. 
In contrast, prior works~\cite{plp,osiris,janus,bmf,triad-nvm} assumed all data remains in NVM (\texttt{NVM-agnostic} application) that may overestimate the overheads of secure NVM ($22.1\times$ slowdown, on average).
However, even with \texttt{NVM-aware} applications, the securing NVM contents can slow down applications by $3\times$, on average, and by up to $22\times$ for memory-intensive workloads. 

Next, we identified that prior studies on secure NVM did not 
consider the limited bandwidth of NVMs.
For example, the write bandwidth of a single Intel Optane NVDIMM is only $2.3$GB/sec~\cite{optane_behavior}.
Prior solutions such as Triad-NVM~\cite{triad-nvm}, ProMT~\cite{promt}, and BMF~\cite{bmf} that execute all security operations in the memory controller cause substantial additional off-chip traffic and bandwidth contention (e.g., $9\times$) to access and persist security metadata.
We find that persisting BMT nodes alone account for up to $71\%$ of the off-chip traffic to NVM.

Driven by these observations, we propose to divide the responsibility of securing NVM contents between the on-chip memory controller and the NVDIMM.
Specifically, we move the \emph{memory-intensive integrity verification} to the NVDIMM to avoid off-chip traffic for updates to BMT nodes while retaining the compute-intensive encryption engine at the memory controller.
Our design, named in-Memory Integrity Verification ({\TOOL}), eliminates the off-chip traffic for updating BMT nodes while conforming to the rigorous threat model~\S\ref{subsec:threat_model} through several novel techniques.

\begin{figure}[]
\centering
     \begin{subfigure}[b]{\linewidth}
         \centering
         \includegraphics[width=\linewidth]{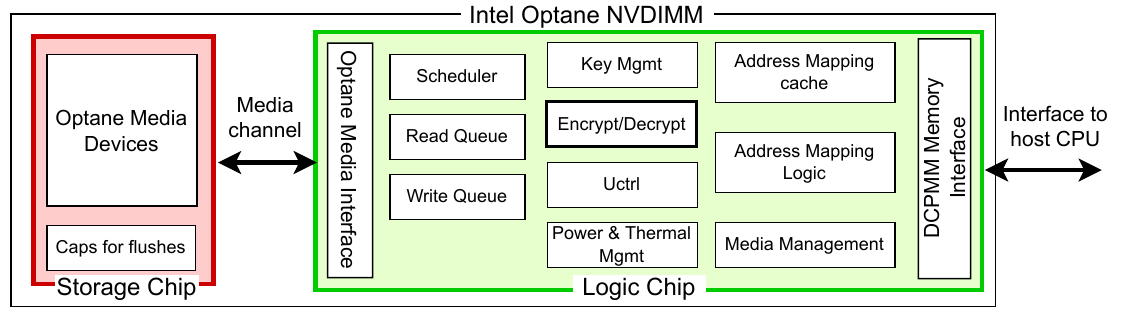} 
     \end{subfigure}
     \hfill
     \vspace{-0.5em}
     \begin{subfigure}[b]{\linewidth}
         \centering
         
         \includegraphics[width=0.8\linewidth]{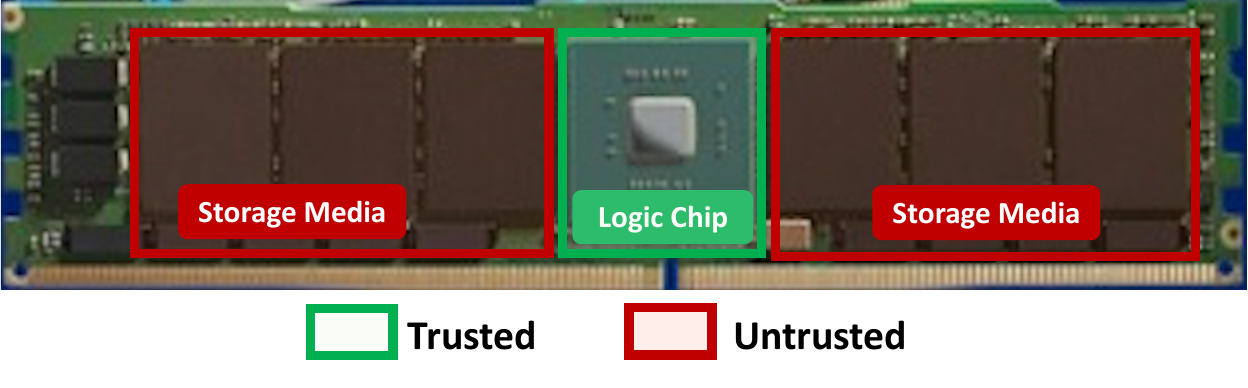}
    \end{subfigure}
    \caption{Image on top is from Intel's Hotchips 2019 presentation that depicts the Optane NVDIMM's internals with an on-DIMM encryption engine\cite{optane-hotchips}. The bottom image shows a real NVDIMM with logic units and untrusted storage media~\cite{optane_chip_diagram}}.
    \label{fig:Optane-dimm}
\end{figure}

An obvious question is whether {\TOOL} is expanding the trusted computing base (TCB) beyond what is the norm today by relying on the logic in NVDIMM for integrity verification.
However, contrary to typical assumptions in research articles, real commercial incarnation of NVDIMM does not consist solely of storage media.
For example, the top part of Figure~\ref{fig:Optane-dimm} shows the internals of commercial Intel Optane NVDIMM~\cite{optane-hotchips}. 
An NVDIMM hosts the logic for remapping physical addresses to media addresses and, more critically, an inbuilt \textit{encryption engine}, besides the NVM media itself~\cite{optane-hotchips}.
The bottom diagram illustrates how commercial NVDIMMs feature distinct storage media and logic chips.
Such a design where the NVMDMM host a significant logic circuitry is not unique to Optane. 
For example, Samsung's new CXL-based memory module Hybrid (CMM-H) that supports NVM functionality also sports compute logic that helps in enforcing memory management polices~\cite{cmm-h}.

Further, on boot, the processor exchanges a secret passphrase with each NVDIMM for authentication~\cite{optane-passphrase}.
Unlike storage media, logic circuitry does not retain states across power cycles that an attacker can exploit. 
Hence, the logic chip on NVDIMM, e.g., in Optane, is trusted, but storage media is not. 
Consequently, {\TOOL} does not extend the TCB beyond what it exits today.
Section~\ref{subsec:threat_model} details our threat model.

Another key advantage of \TOOL{} is that it \emph{scales with the per-DIMM capacity} rather than the aggregate NVM capacity in a system. 
Since the height of BMT is proportional to the amount of data it protects, as NVM capacity increases, so does BMT's height, resulting in increased overheads.
In {\TOOL}, each NVDIMM maintains a separate (sub-)BMT whose height is determined solely by that DIMM's capacity and is independent of the aggregate NVM capacity in the system. 

Furthermore, {\TOOL} employs a novel \emph{split BMT cache} that uses the fact that BMT nodes near the root (upper-level BMT nodes) are updated more frequently than those further down (lower-level nodes). 
We, thus, partition a typical BMT cache into two. 
One component is dedicated to buffering only the upper-level nodes.
With only a few upper-level nodes, a 72-entry buffer is adequate to avoid replacements. On a power-down event, the contents of this buffer is persisted. 
The second part is the usual BMT cache that caches only the lower-level nodes and is non-persistent.
This reduces writes to NVM media with limited write endurance.

We evaluated {\TOOL} with 10 workloads, each with two versions: \texttt{NVM-aware} and \texttt{NVM-agnostic}.
We demonstrated that for \texttt{NVM-aware} workloads, {\TOOL} enables strong security guarantees with fine-grain verifiability at a performance overhead of only $55\%$ over a system with no security guarantees.
We quantitatively compare {\TOOL} it against many prior works -- Triad-NVM~\cite{triad-nvm}, PLP~\cite{plp}, ProMT~\cite{promt}, SBMF~\cite{bmf}. 
Our results for \texttt{NVM-aware} applications show that {\TOOL} achieves average speedup of $2\times$, $1.78\times$, $1.68\times$, $1.59\times$ and $1.38\times$ over a baseline with all security guarantees but no optimizations, \texttt{PLP}, \texttt{Triad}, \texttt{ProMT}, and \texttt{SBMF}, respectively.

\noindent In summary, we make the following contributions:
\begin{itemize}[leftmargin=*]
\item We quantitatively demonstrate that under realistic off-chip NVM bandwidth, persisting BMT nodes emerges as the key  bottleneck for secure NVM with fine-grain verifiability. 
\item We designed {\TOOL}, that leverages in-memory computing on NVMDIMM for integrity verification of persistent data. 
\item We show that, in comparison to prior techniques, \TOOL{} provides strong security guarantees at lower overheads (\S\ref{sec:evaluation}).
\end{itemize}

%% file: PaperContent/3_background.tex
\section{Background}
\label{sec:background}

\medskip

\noindent\textbf{Non-volatile Memory:}
Non-volatile Memory (NVM) technologies enable byte-addressable loads and stores to persistent data. 
NVM enables fine-grain persistence at latencies \textit{comparable} to volatile DRAM~\cite{yang:fast:2020}.
Intel's Optane DC Memory was the first commercial NVM.
While Intel recently discontinued Optane, Samsung's new memory-semantic SSD technology (MS-SSD) and CXL-Memory module Hybrid (CMM-H) support persistent memory mode~\cite{mem-ssd}.
More alternative NVM technologies such as 3D flash memory~\cite{kioxia-flash} and Everspin's STT-RAM~\cite{everspin-stt-ram} are also emerging.
Further, Compute Express Link (CXL)~\cite{cxl}, the emerging industry-wide standard for disaggregated computing, has incorporated a global persistent flush operation for CXL-attached PM~\cite{gpflush:cxl}. 
It suggests the industry's expectation that many vendors are likely to offer NVM products in the future~\cite{cxl-pmem}. 
In short, while Optane was the first commercial NVM, it is unlikely to be the last.

\medskip

\noindent\textbf{Security challenges and techniques in NVM:} Unlike DRAMs, NVMs are susceptible to data remanence attacks as the data persists across power cycles. 
An attacker may passively read or actively manipulate the data using splicing, spoofing, and replay attacks~\cite{attacks}.
In a splicing attack, an attacker replaces a memory block with one at a different location.
In spoofing, the attacker substitutes a malicious memory block for an existing one. 
A replay attack replaces a memory block at a given address with older data.
Next, we discuss typical techniques employed to defend against common security challenges in NVM. 

\medskip

\begin{figure}[]
\centering
\includegraphics[width=\linewidth]{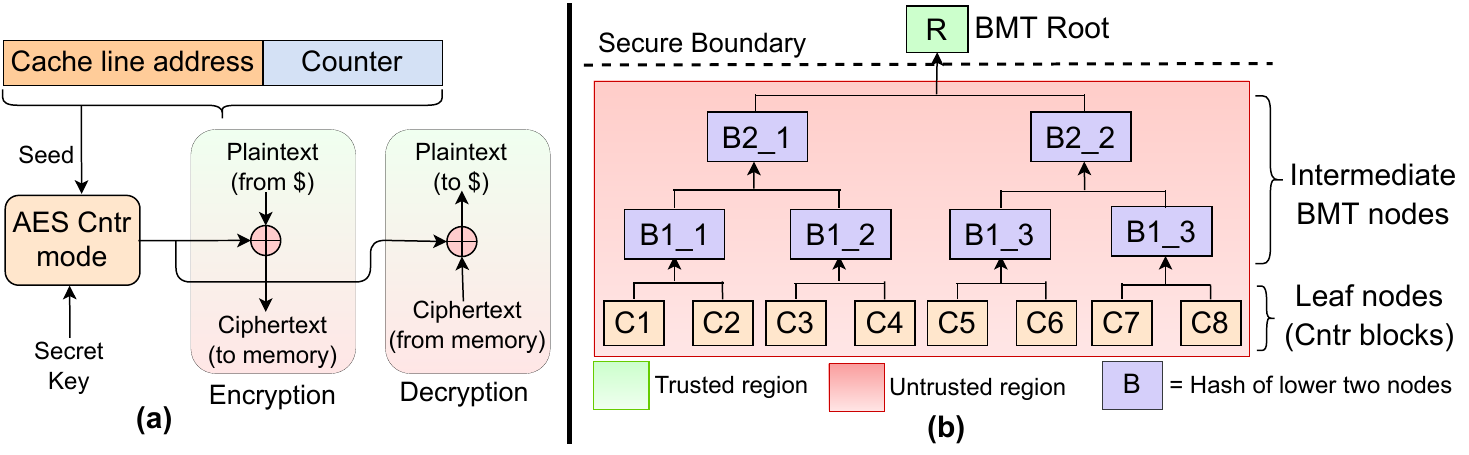}
\caption{(a) Counter-mode encryption and (b) 2-array BMT}\label{fig:BMT}
\end{figure}

\noindent \textbf{\circled{\textbf{1}} Confidentiality via Counter-Mode Encryption (CME).}
Data confidentiality is achieved by encrypting cache blocks upon eviction from CPU caches (trusted) to the NVM. 
It is common to use split storage CME~\cite{selective_atomicity} for low decryption latency. 
In CME, each cache line is associated with a counter that is incremented upon each cache eviction, along with a secure key to generate a One-Time Pad (OTP). Data is encrypted (decrypted) by XORing the generated OTP with the plaintext (ciphertext) as shown in Figure~\ref{fig:BMT}(a).

\medskip

\noindent \textbf{\circled{\textbf{2}} Data and counter integrity via SMAC.}
One must also verify that the data read from NVM is not tampered with~\cite{mem_authentication}. 
Stateful Message Authentication Codes (SMACs) detect data integrity violations by computing a hash over ciphertext, CME's counter, the address, and a private key. 
Modifications to any of these result in a hash mismatch, detecting spoofing and splicing attacks. 
SMACs could detect replay attacks if they are stored within the trusted CPU. 
However, SMACs are stored on untrusted NVM as they need large space.

\medskip
\noindent \textbf{\circled{\textbf{3}} Counter freshness via Bonsai Merkle Tree (BMT).}
BMT~\cite{bmt} is a hash tree for detecting stale counters to thwart replay attacks.
It recursively computes a hash over blocks storing CME counters and stores them in the tree nodes (Figure~\ref{fig:BMT}(b)).
BMT root is stored securely within the CPU, while the intermediate nodes and counters reside on the NVM.
Counters retrieved from NVM are verified by computing the hash tree and comparing it with the BMT root stored in the CPU. 
BMT, with SMACs, ensures the integrity and freshness of data and counters, stopping replay attacks.

\medskip
\noindent \textbf{Security metadata caching.}
Security metadata is typically cached in on-chip metadata caches to limit overheads of fetching the metadata from off-chip NVM.
For example, when a counter's ancestor node (in BMT) is cached, no further verification from that ancestor node to root is required as its integrity has already been verified when it was earlier fetched into the metadata cache. This decreases the verification overhead of reading a counter.
Likewise, counters and SMACs are cached in the trusted CPU.

%% file: PaperContent/4_Security_guarantees.tex
\section{NVM security guarantees}
\label{sec:security_guarantees}
\begin{figure}[]
\centering
    \includegraphics[width=.85\linewidth]{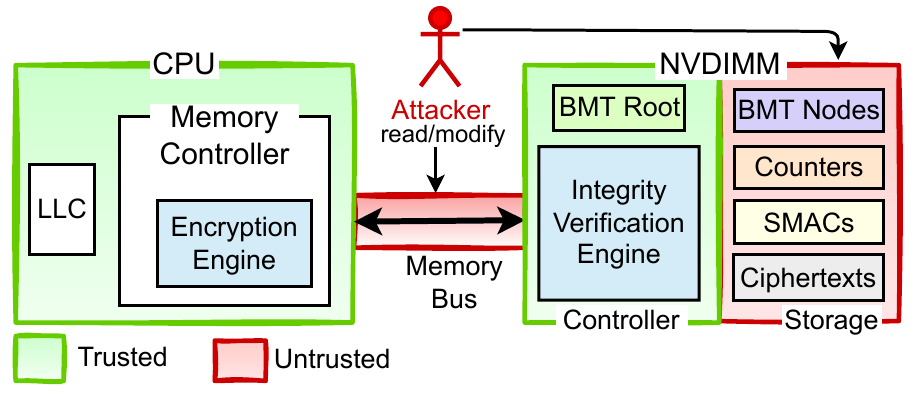}
    \caption{Threat Model and TCB under consideration.}
    \label{fig:Threat-model}
\end{figure}

\begin{scriptsize}
\begin{table}[t]
    \scriptsize
    \centering
    \caption{Security guarantees and associated security tuple.}
    \label{table:security_guarantees}    
    \begin{tabular}{p{4.3cm}|p{3.5cm}}
        \hline
        \textbf{Security guarantees} & \textbf{Security tuple to be persisted} \\
        \hline
        \hline
        Confidentiality & \emph{Ciphertext, \textbf{Counter}}\\ 
        \hline
        $+$ Data \& Counter integrity & \textit{$+$ \textbf{SMAC}} \\
        \hline
        $+$ Freshness of Data and Counter & \textit{$+$ \textbf{BMT root}}  \\
        \hline
        $+$ Fine-grain unverifiable region detection & \textit{$+$ \textbf{Intermediate BMT nodes}} \\
        \hline
    \end{tabular}
\end{table}
\end{scriptsize}

\begin{scriptsize}
\begin{table*}[t]
    \scriptsize
    \centering
    \caption{Classification of prior works based on security guarantees they provide and their ability to detect unverifiable region.}
    \label{table:prior_works}
    \begin{tabular} 
{|p{1.8cm}|C{.93cm}|C{.8cm}|C{.99cm}|p{6.6cm}|C{1cm}|C{1.2cm}|C{1.2cm}|}
    \hline
    \textbf{Prior works} & \textbf{Data \& counter integrity} & \textbf{Fine-grain verification} & \textbf{Recovery time} & \textbf{Security tuple} & \textbf{Security tuple persist ordering} & \textbf{In-memory integrity verification} & \textbf{Factors in limited NVM bandwidth }\\
    \hline
    \textbf{SCA}~\cite{selective_atomicity} & \xmark & \xmark & - & \textit{Ciphertext, Counter} & \xmark & \xmark & \xmark\\ 
    \hline
    \textbf{SecPM}~\cite{secpm} & \xmark & \xmark & - & \textit{Ciphertext, Counter} & \xmark & \xmark & \xmark\\ 
    \hline        
    \textbf{Osiris}~\cite{osiris} & \cmark & \xmark & $hr$-scale & \textit{Ciphertext, Counter, SMAC, BMT (root)} & \xmark & \xmark & \xmark \\
    \hline
    \textbf{Anubis}~\cite{anubis} & \cmark & \xmark & $ms$-scale & \textit{Ciphertext, Counter, SMAC, BMT (root)} & \xmark & \xmark & \xmark \\
    \hline
    \textbf{Janus}~\cite{janus} & \cmark & \xmark & $hr$-scale & \textit{Ciphertext, Counter, SMAC, BMT (root)} & \xmark & \xmark & \xmark \\
    \hline
    \textbf{Triad-NVM}~\cite{triad-nvm} & \cmark & \textbf{partial} & $ms$-scale & \textit{Ciphertext, Counter, SMAC, BMT (root, lower intermediate nodes)} & \xmark & \xmark & \xmark \\
    \hline
    \textbf{PLP}~\cite{plp} & \cmark & \xmark & $hr$-scale & \textit{Ciphertext, Counter, SMAC, BMT (root, intermediate nodes)} & \cmark & \xmark & \xmark \\
    \hline        
    \textbf{Shield-NVM}~\cite{shield-nvm} & \cmark & \cmark & $ms$-scale & \textit{Ciphertext, Counter, SMAC, BMT (root, intermediate nodes)} & \xmark & \xmark & \xmark\\
    \hline
    \textbf{BMF}~\cite{bmf} & \cmark & \xmark & $min$-scale & \textit{Ciphertext, Counter, SMAC, BMT (root)} & \cmark & \xmark & \xmark\\
    \hline
    \textbf{ProMT}~\cite{promt} & \cmark & \cmark & $\mu$s-scale & \textit{Ciphertext, Counter, SMAC, BMT (root, intermediate nodes)} & \cmark & \xmark & \xmark\\
    \hline
    \hline
    \textbf{\TOOL} & \cmark & \cmark & $\mu$s-scale & \textit{Ciphertext, Counter, SMAC, BMT (root, intermediate nodes)} & \cmark & \cmark & \cmark\\
    \hline
    \end{tabular}
\end{table*}
\end{scriptsize}

\subsection{\textbf{Threat model and trusted computing base}}
\label{subsec:threat_model}
The primary security challenge with NVDIMM is data remanence, which leaves data vulnerable across power cycles. 
We assume a strict threat model where an adversary can scan, snoop and modify data on the memory bus, replay earlier memory operations, and modify data on NVM media.

The system's trusted computing base (TCB) consists of the SoC and an integrity verification engine (IVE) logic on-board NVDIMMs (Figure~\ref{fig:Threat-model}). 
The memory bus and NVM storage media are untrusted.
The trusted boundary within NVDIMM ends at the interface between IVE and storage media.
This TCB is based on real deployment of NVM, such as Intel's Optane NVDIMM (Figure~\ref{fig:Optane-dimm}), which trusts the SoC and onboard circuitry for security operations such as authentication and encryption but not the storage media.
While Optane's threat model considers the memory bus as trusted, we consider a stricter threat model where the memory bus is untrusted.

\subsection{\textbf{Security guarantees and associated constraints}} 
\label{subsec:security_metadata}
Guaranteeing security, verifiability, and recovery guarantees require persisting various security metadata, referred to as security tuples (STs), along with the data.
This presents a trade-off between performance and security guarantees. 
We discuss different security guarantees and the associated security tuples (summarized in Table~\ref{table:security_guarantees}) and also discuss related prior works in that context. Table~\ref{table:prior_works} highlights various security, verifiability, and recoverability properties of prior works, and {\TOOL}.

\medskip
\noindent \textbf{\circled{1} Confidentiality:} 
For CME, the confidentiality of the data in the NVM can be guaranteed with the security tuple consisting of the ciphertext and the counter. 
All prior work on secure NVM provides this minimum guarantee.

\medskip
\noindent \textbf{\circled{2} Data and counter tampering detection:}
When the security tuple comprising ciphertext and counter is extended with SMAC, one can additionally detect data tampering attacks such as spoofing and splicing.
Any modifications to ciphertext, counter, or SMAC can be detected by recomputing the SMAC within TCB and matching it with the stored SMAC. 
As seen in Table~\ref{table:prior_works}, many prior works also enable this level of security.

\medskip
\noindent \textbf{\circled{3} Guarding against replay attacks:}
Detecting data and counter tampering is not enough to protect against replay attacks.
An attacker can snoop and feed older responses to the SoC.
The BMT over the counters can detect staleness of the counters by storing the BMT root in the TCB.
Thus, when the security tuple is extended to cover BMT root, replay attacks can be detected too. Triad-NVM~\cite{triad-nvm}, PLP~\cite{plp}, BMF~\cite{bmf} and ProMT\cite{promt} guards against replay attack. 

\medskip
\noindent \textbf{\circled{4} Fine-grain verifiability and fast recovery:}
If the intermediate BMT nodes are not persisted, then on a replay attack it is possible that all SMACs match their corresponding ciphertext and counter, but the reconstructed BMT root node differs from that in the TCB. 
Even if a single 64-byte block data or a single 8-byte counter is tampered with, entire TBs of NVM would become unverifiable\cite{triad-nvm}. 
Also, not persisting the intermediate nodes result in recovery times in many hours~\cite{anubis}. 
Thus, for fine-grain verifiability and fast recoverability, ST must also include intermediate BMT nodes.
However, most prior works avoid persisting intermediate nodes or persist only parts of the BMT (e.g., Triad-NVM~\cite{triad-nvm}) to reduce overheads.

\subsection{\textbf{Crash consistency in secure NVM}} 
\label{subsec:crash_consistency}
The fundamental purpose of the persistence offered by NVM is to enable recoverability. Ensuring both the recoverability and security of data on NVM adds additional constraints on when and how the security tuple is persisted.

\medskip
\noindent \textbf{\circled{1} Persisting security tuple with data:} Security tuple defined in accordance with the required security guarantees should be persisted along with the data to facilitate recovery. If any of the tuple elements are not updated
along with data, then integrity, verification fails or results in incorrect plaintext~\cite{plp, selective_atomicity, triad-nvm}.

\medskip
\noindent \textbf{\circled{2} Ordering between persisting multiple security tuples:} To ensure recovery after a crash, security tuples must be updated in the same sequence as data~\cite{plp}. If not, it results in either BMT or MAC verification failure or in incorrect plaintext.
Further, updates to BMT from leaf to root must also follow the data persist order. 
Otherwise, BMT will be out of sync with data updates after a system crash, preventing correct data recovery. However, most prior works neglected this need. 

From Table~\ref{table:prior_works}, we notice that only two prior works, Shield-NVM~\cite{shield-nvm} and ProMT~\cite{promt}, offer many desirable security, verifiability, and recoverability guarantees. 
Shield-NVM relied on SMAC to recover older security tuple to the newest version. 
However, it fails if the attacker modifies the SMAC itself. 
Further, Shield-NVM's epoch-based model did not ensure security tuple persist ordering.
While ProMT~\cite{promt} persists intermediate BMT nodes to either hotMT or globalMT depending on the classification of the page based on access frequency.
However, it ignores the impact on the limited NVM bandwidth (analysis in \S\ref{sec:evaluation}).
The \TOOL{} is the only work that factors limited NVM bandwidth and leverages in-memory integrity verification. 

%% file: PaperContent/5_motivation.tex
\section{Bottleneck analysis, key insights and ideas}
\label{sec:motivation}

\begin{figure}[]
\centering

\includegraphics[width=.8\linewidth]{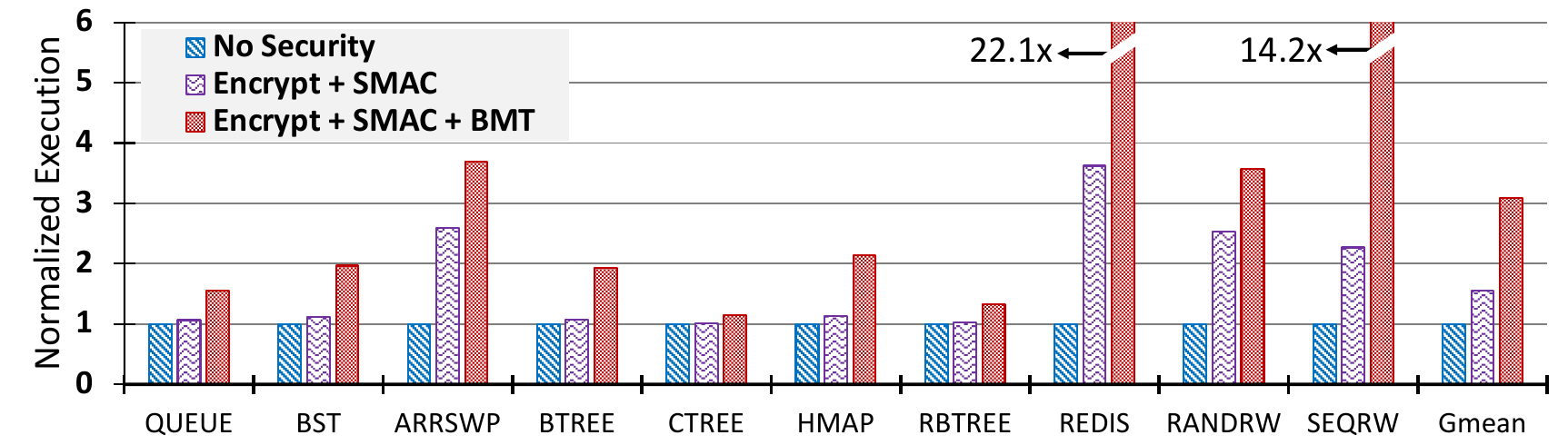}
\caption{Impact of security metadata computation and persistence on NVM-aware workloads.} 
\label{fig:motivation}
\end{figure}

\begin{figure}[]
\centering
\includegraphics[width=.8\linewidth]{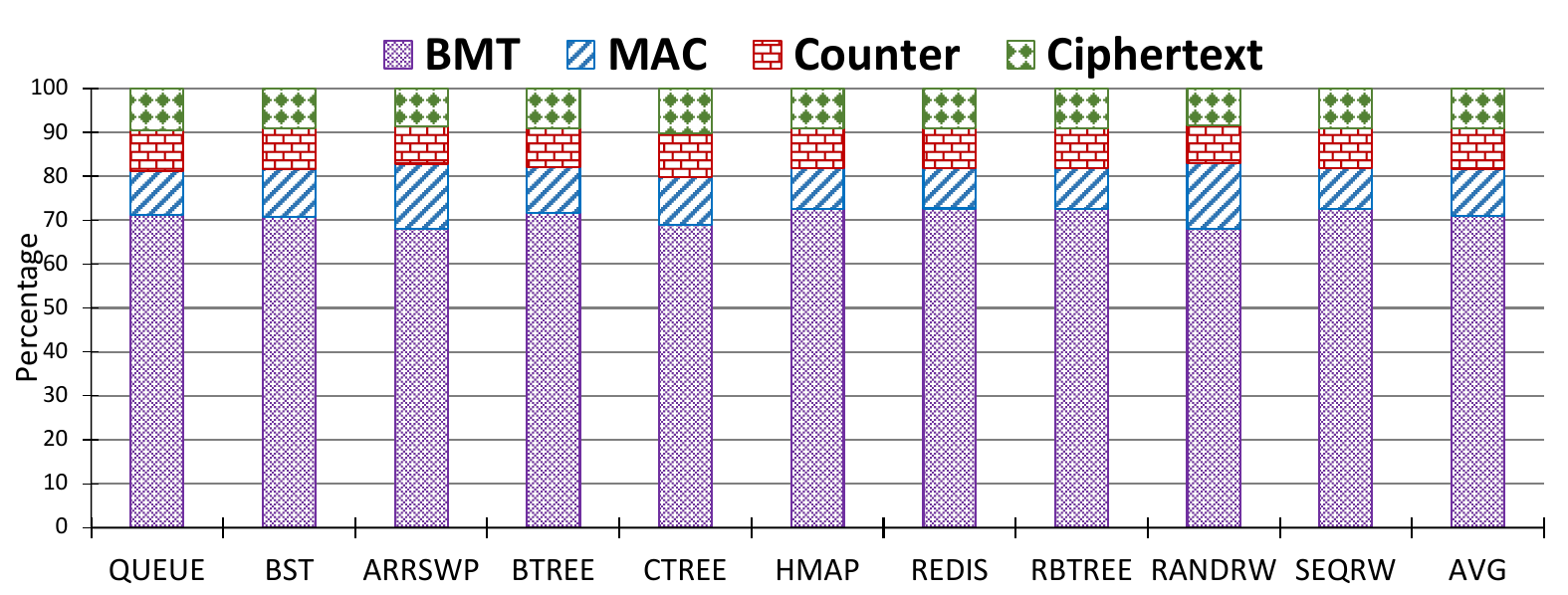}
\caption{Contribution of each element of security tuple to off-chip data movement.} 
\label{fig:data_movement_analysis}
\end{figure}

\begin{figure*}[h!]
\centering
\includegraphics[width=0.7\linewidth]{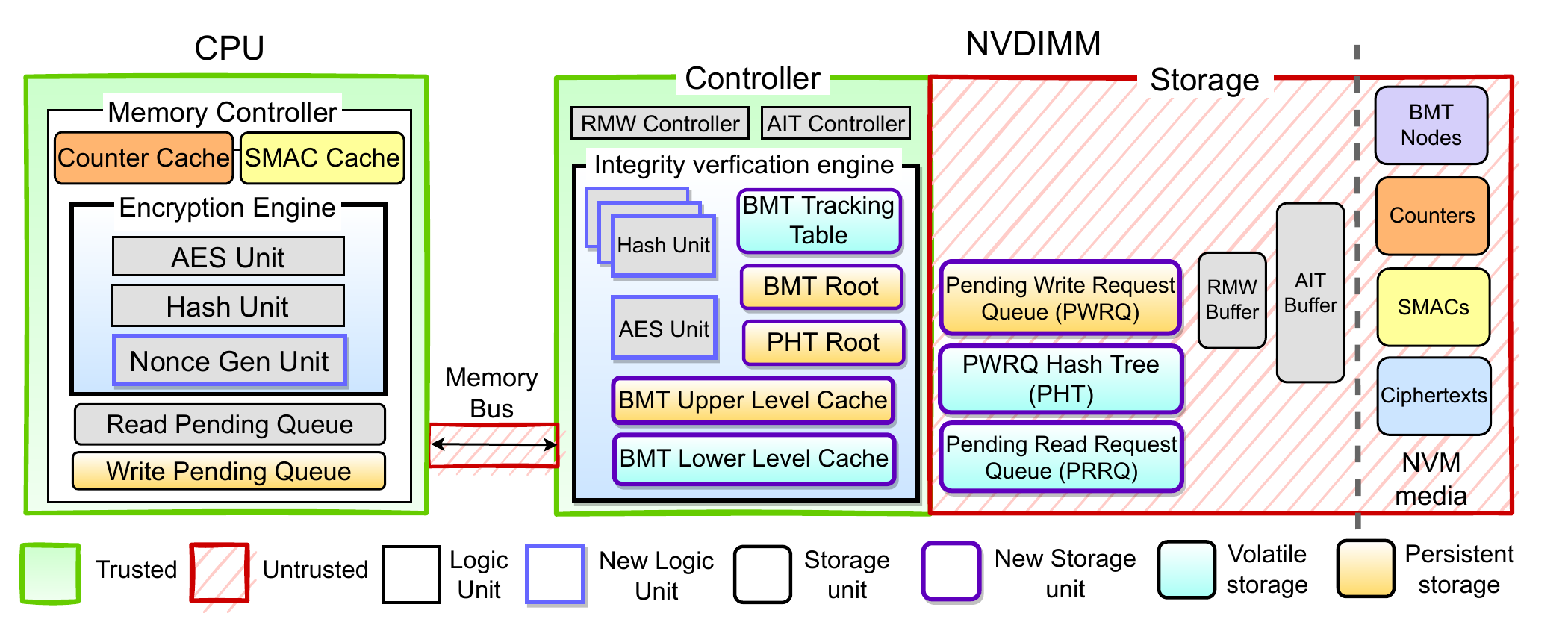}
\caption{Architecture of {\TOOL} with encryption engine at memory-controller and integrity verification engine at NVDIMM.}\label{fig:architecture_design}
\end{figure*}

Figure~\ref{fig:motivation} quantifies the performance overheads of ensuring security guarantees for various applications (\S\ref{sec:evaluation} details methodology).
We use \texttt{NVM-aware} workloads that keep only the data needed for correct recovery on the NVM~\cite{whisper} while the rest of the data is in volatile memory.
This helps estimate realistic overheads of secure NVM.
We also model security metadata caches, such as BMT node cache, counter cache, and SMAC cache, which reduces overheads (detailed in \S\ref{sec:evaluation}).

The first bar in Figure~\ref{fig:motivation} represents the execution time when no security guarantee is provided (\texttt{No security}). 
The next bar (\texttt{Encrypt+SMAC}) represents the execution time when encryption and SMAC computation are performed, along with persisting  \emph{counter} and \emph{SMAC}, but without BMT
(this cannot detect replay attacks). 
The third bar (\texttt{Encrypt+SMAC+BMT}) represents the system where all three security operations are performed (CME, SMAC, and persisting all BMT nodes).
It provides all security guarantees along with fine-grain verifiability and can detect replay attacks. All plot is normalized to that of \texttt{No security} bar.

From Figure~\ref{fig:motivation}, we notice that even under realistic scenario of \texttt{NVM-aware} workloads where only a fraction of stores directed to NVM~\cite{whisper}, achieving all the security and verifiability properties described before (\S\ref{sec:security_guarantees}) still incurs significant performance overheads of $3\times$ on an average and up to $22.1\times$ in worst case.
Importantly, persisting BMT nodes contribute significantly to the overheads. 
While prior works have ignored the constrained bandwidth of commercial NVDIMMs ($\sim2.3$GB/sec~\cite{optane_behavior} for Optane), off-chip data movement in secure NVM is a key factor behind the overheads.

Figure~\ref{fig:data_movement_analysis} shows how different components of the security tuple contribute to off-chip data movement. 
We notice that often $>70\%$ of off-chip data movement can be attributed to BMT node transfer alone. 

In summary, we observe that even for realistic execution scenarios with \texttt{NVM-aware} applications, secure NVM can incur significant overheads, although typically less than as claimed in prior works.
Importantly, the overhead of persisting  BMT nodes is the key bottleneck that incurs most of the off-chip data movement to bandwidth-constrained NVMs.

%% file: PaperContent/6_Key_Ideas.tex
\subsection{\textbf{Key ideas}}
\label{sec:key_ideas}
Our goal is to achieve strong security and verifiability guarantees ( \S\ref{sec:security_guarantees}) while limiting the cost of BMT updates by leveraging the following ideas.

\medskip
\noindent \textbf{\circled{1} In-memory integrity verification.} BMT updates are memory-intensive operations. 
Hence, we offload the computation of BMT nodes to the NVDIMM's integrity verification engine (IVE). This avoids off-chip data movement that ails fine-grain integrity verification.
We leverage the insight that real NVDIMMs already contain trusted logic units, separate from the untrusted NVM media. 

\medskip

\noindent \textbf{\circled{2} Scaling integrity verification.} 
A key advantage of placing the IVE on the NVDIMM is that it scales with the per-DIMM capacity instead of aggregate NVM capacity.
A BMT's height is proportional to the memory capacity it secures.
In our design, each NVDIMM maintains independent BMT.
Thus, the aggregate capacity of the NVM in a system can be scaled by attaching multiple NVDIMMs without increasing the height of BMTs, hence, the cost of integrity verification.

\medskip

\noindent \textbf{\circled{3} Limiting updates to top-level BMT nodes.} 
The top-level BMT nodes (near the root) are updated more frequently than lower-level BMT nodes due to the tree structure. Thus, we split the typical BMT cache into two parts.
One part is dedicated to buffering only the few upper-level nodes. Since there are only a few upper-level nodes, a small buffer (e.g., 64 entries) avoids replacement. 
The buffer contents are persisted only at power down, avoiding large number of writes to these nodes. 
The second part is the typical BMT cache (volatile) that caches lower-level BMT nodes and replaces items as needed.

%% file: PaperContent/7_design.tex
\section{Design and Implementation of \TOOL{} }
\label{sec:design}
The proposed {\TOOL} has two key components that are responsible for computing and maintaining security metadata: the first, as with previous works~\cite{triad-nvm,janus,bmf,promt}, is an encryption engine (EE) hosted on the memory controller (MC) and the second is a per-NVDIMM integrity verification engine (IVE), which is the new key hardware component. 

Figure~\ref{fig:architecture_design} depicts the {\TOOL}'s architecture that builds upon Optane NVDIMM's design.
While Optane is discontinued for commercial reasons, it serves as a realistic baseline design in the absence of deeper technical details of alternatives such as Samsung's MS-SSD in the public domain. 

Figure~\ref{fig:architecture_design} marks newly added components, trusted and untrusted components, and those which are volatile and non-volatile in different colors/patterns. 
EE on the CPU is responsible for CME and SMAC as in today's design~\cite{triad-nvm,janus,osiris}. 
We make only minor enhancements to the EE by adding a hardware nonce unit (random number generator) to help detect replay attacks on the memory bus (detailed later).
The memory controller on the CPU also has Write Pending Queue (WPQ)~\cite{adr}, as well as Read Pending Queue (RPQ) as in a typical design today.
As in the Xeon CPUs that support Optane NVM, WPQs are part of a non-volatile domain since it guarantees that the contents of WPQ will be flushed to NVM at a power-down event. 
We further extend each WPQ entry to collect the security tuple corresponding to the data being written.

The IVE is responsible for the counters' integrity and fine-grain verifiability, and its design is the key contribution of this work.
It hosts several new queues and logic units (e.g., AES unit).
We will detail the role of each component while discussing the working of \TOOL{} later in this section. 
Similar to contemporary designs~\cite{osiris,triad-nvm}, {\TOOL} stores frequently used security tuples in the metadata caches – the counter and SMAC caches in EE and the BMT buffers/caches in IVE. {\TOOL} ensures crash-consistent security by storing data blocks and their associated security tuples to the NVDIMM with each data persistence.
Metadata caches are write-through. 
The IVE on the NVDIMM enables fine-grain counter verification without needing off-chip accesses for BMT nodes while utilizing the NVDIMM's \textit{internal} bandwidth.
By retaining EE in the memory controller, {\TOOL} ensures security even when the memory bus is untrusted. 
We now detail the working of each component along with their purpose.

\subsection{\textbf{Encryption Engine (EE)}}
\label{subsec:EE} 
When the CPU sends a read/write to the memory controller on an LLC miss or explicit cache line flush, the EE is activated. We explain in detail how read and write requests are performed when they hit and miss in the metadata caches.

\medskip
\noindent\textbf{Read requests}: EE examines the counter and SMAC caches upon a read request and reacts as follows. 

\medskip
\noindent \textit{\textbf{Case \circled{\small{A}}} Hits in both Counter \emph{and} SMAC cache:}
EE retrieves only the ciphertext from NVDIMM and computes the SMAC on the ciphertext and compares it to the SMAC stored in the SMAC cache. Following the SMAC verification, the plaintext is obtained by XORing the OTP generated by the counter (from the counter cache) and the ciphertext.

\medskip
\noindent \textit{\textbf{Case \circled{\small{B}}} Miss in either counter \emph{or} SMAC cache:} The missed metadata is fetched from the NVDIMM, besides the ciphertext. SMAC is recalculated after receiving the responses and then verified against the SMAC retrieved from either SMAC cache or received from NVM.

\medskip
\noindent \textit{\textbf{Case \circled{\small{C}}} Misses in both counter \emph{and} SMAC cache:} 
When ciphertext, counter, and SMAC values are all obtained from NVM, there is no way to determine whether the values received at the memory controller have been tampered by a replay attacker snooping on the memory bus. 
This challenge is specific to {\TOOL} since the BMT root does not reside in the memory controller.
In traditional design, the BMT root would have helped detect replay attack on the memory bus.
We address this challenge by introducing a \textbf{nonce} (random number) generator in EE (Figure~\ref{fig:architecture_design}).
The read request for a counter is augmented with a nonce. 
While sending back the verified counter, NVDIMM encrypts the counter using the nonce received along with the request.
On receiving the encrypted counter, the memory controller decrypts it using the corresponding nonce. 
We explain how this helps ensure security guarantees in more detail in \S\ref{sec:security_discussion}.

\medskip
\noindent \textbf{Write requests}: Before transmitting cache blocks off-chip to the NVDIMM, EE reacts as follows.

\medskip
\noindent\textit{\textbf{Case \circled{\small{A}}} Counter cache \emph{and} SMAC cache hit:}
EE immediately produces the ciphertext using CME. Additionally, EE computes and saves the SMAC in the SMAC cache, and populates the WPQ with the ciphertext and SMAC as part of the security tuple. 
In \TOOL{}, EE is not required to send the updated counter, as the IVE on the NVDIMM controller increments the counter on receiving the ciphertext writeback.

\medskip
\noindent\textit{\textbf{Case \circled{\small{B}}} SMAC cache hit but counter cache miss:}
EE creates the fetch request for the counter and the ciphertext and sends them to the NVDIMM.
The most recently stored ciphertext must be retrieved from the media, as this is required to recalculate the SMAC upon arrival at the EE and verify the counters' freshness via SMAC comparison. 
Following SMAC verification, the counter is incremented and used to generate the OTP necessary to obtain the new ciphertext.

\medskip
\noindent\textit{\textbf{Case \circled{\small{C}}} Misses in both counter and SMAC cache:} 
As with the ciphertext read, a unique case arises on misses to both the counter cache and the SMAC cache for {\TOOL}. At EE, we solve this problem through the use of a nonce generator, similar to the read request (detailed in \S\ref{sec:security_discussion}).

\subsection{\textbf{Integrity verification engine (IVE) on NVDIMM}}
\label{subsec:IVE} 
\medskip
Once NVDIMM receives a read or write request, it is queued in the Pending Read Request Queue (PRRQ) or Pending Write Request Queue (PWRQ)).
The IVE ensures the integrity of every request made to the NVDIMM.

\medskip
\noindent\textbf{Read requests:} PRRQ is a 16-entry (default) queue that receives read requests for ciphertext, counter, and SMAC. 
IVE retrieves the requested items from NVDIMM's internal buffers or the media and sends them to the memory controller.
However, for counter read requests, it must first verify its integrity as follows.
The IVE retrieves the counter block from the NVM's internal buffers or the NVM media.
IVE then runs the typical integrity check using BMT calculation.
It searches the BMT caches and buffers (described below) and compares them to the BMT root.
Once the counter is verified, it is sent to the memory controller. In the particular case when there is a miss in both the counter cache and SMAC cache in the memory controller, EE sends a nonce with a counter read request.
IVE, in such cases, uses the received nonce to encrypt the counter before responding.
This helps guard against replay attacks on the memory bus. 

\medskip
\noindent\textbf{Write requests:} 
The incoming ciphertext and its SMAC are inserted into the PWRQ.
The other security tuple members, such as counter and BMT nodes, are calculated and placed into the PWRQ.
Then, the corresponding counter is fetched from the internal buffers or the media.
Note that an NVDIMM does not have a counter cache. 
We find Optane's existing internal buffers, such as RMW and AIT, are enough for caching counter blocks. 
The IVE verifies the freshness of the counter read from the untrusted media by computing the BMT nodes and finally comparing them with the BMT root securely stored on the IVE. 
Once verified, the counter is incremented to account for the new store to a data cache line. The counter is then added to the PWRQ entry, and the BMT is updated from leaf to root. The updated intermediate BMT nodes are added to the PWRQ. 
The IVE calculates the SMAC using the counter, the plaintext, and the address. The SMAC is compared to the one received with the ciphertext to detect tampering on the memory bus.
Once the entire security tuple for a ciphertext is ready in PWRQ, PWRQ entry is drained to the media.

\begin{figure*}[h!]
\centering
\includegraphics[width=0.8\linewidth]{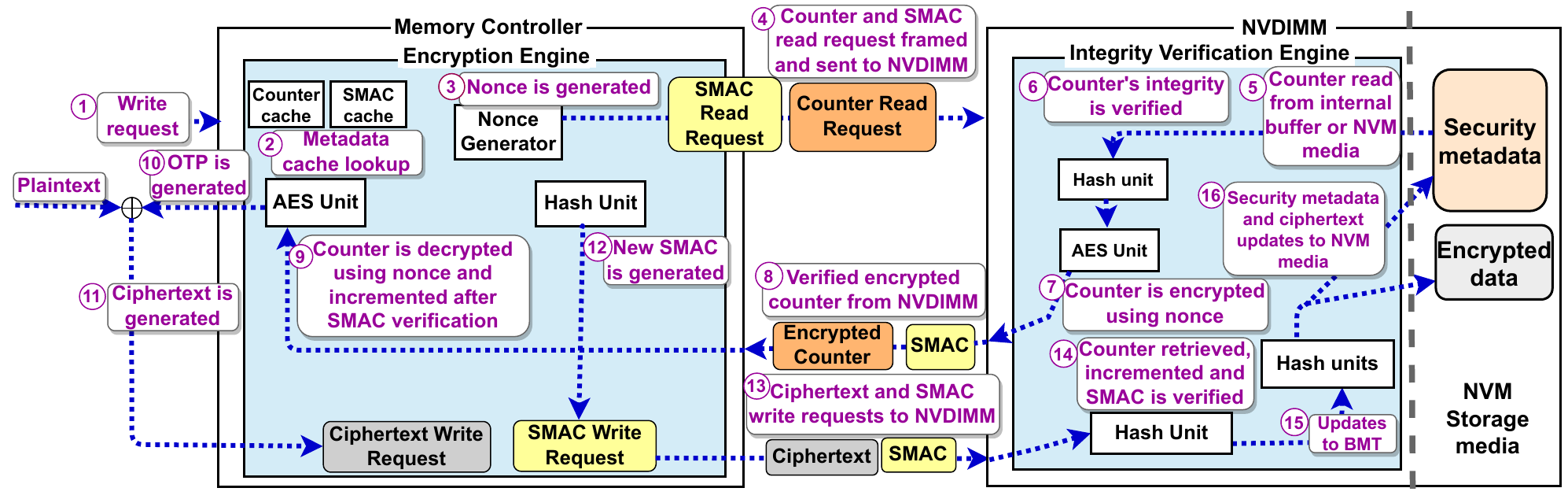}
\caption{\label{fig:Store-flow}Detailed end-to-end flow of a write request.}
\end{figure*}

\subsection{\textbf{Efficient BMT updates}}
\label{para:BMTupdates} 
As persisting BMT nodes is the key performance bottleneck, we propose two BMT update optimizations.

\medskip
\noindent \textbf{Split cache technique:} The upper-level BMT nodes are updated more frequently than the lower-level nodes (near leaf) due to its tree structure. We leverage this by partitioning the traditional BMT cache into two. 
The BMT upper-level cache (Figure~\ref{fig:architecture_design}) is a small buffer to hold the uppermost levels (top 3) BMT nodes \textit{without} replacement. 
This is possible since there are a small number of upper-level nodes.
The BMT lower-level cache is a traditional BMT cache but is used to hold only lower-level nodes. Segregating upper-level and lower-level entries ensures that lower-level entries cannot replace more useful upper-level entries.
Importantly, frequent updates to the BMT upper-level nodes do not cause frequent writes to NVM media since the contents of the upper-level cache are written to NVM media only at a power down event.  

\medskip
\noindent \textbf{Pipelined updates:} The IVE updates BMT in a pipelined manner~\cite{plp} from the leaf to the root. 
It utilizes the same number of hash units as the BMT's level count (here, 9 for a 512GB NVDIMM). Multiple BMT updates (maximum 9) could be running concurrently at any moment in time. However, ensuring consistency requires the updates to happen in the same sequence as the persisting of the ciphertext to the media. Similar to PLP's PTT~\cite{plp}, we employ the BMT tracking table for this purpose which holds a pointer to the PWRQ entry undergoing BMT updates.


\input{PaperContent/13_store_flow}

\section{Attack mitigation, recovery \& cost analysis}
\label{sec:security_discussion}

\subsection{\textbf{Security analysis and attacks mitigation}}

The 64-byte BMT root within IVE is deemed secure. 
To prevent manipulation, one can retain multiple copies of the BMT root, and a read to the BMT root would be valid if all the copies matches.
To protect PWRQ entries (default, 64) from malicious modifications, we use shallow (three-level), eight-array integrity tree, named PWRQ Hash Tree (PHT). 
The PHT root is securely maintained to detect any potential tampering with PWRQ entries. 
We next analyze how \TOOL{} mitigates various attacks mentioned in \S\ref{sec:security_guarantees}.

\medskip
\noindent\textbf{Confidentiality, spoofing and splicing attack:} The CME in the memory controller guarantees confidentiality.
Data outside the SoC is always encrypted. 
By computing and verifying SMAC at both ends of the memory bus, tampering with the data is detected thus mitigating spoofing and splicing attacks. 

\medskip
\noindent\textbf{Replay attack:}
Replay attacks requires a deeper attention in our design.
If the SoC has a hold of either a counter, ciphertext, or SMAC, then any tampering of any response sent by the NVDIMM on the memory bus is detected as before (\S\ref{subsec:security_metadata}). 
However, if none are present in the SoC, then attacker may possibly replay older counter, SMAC and the corresponding ciphertext.
Unlike in traditional designs, in the absence of the BMT root on the SoC in our design, one may not be able to detect such tampering on the memory bus.

{\TOOL} guards against this by incorporating a nonce (random number) generator (Figure~\ref{fig:architecture_design}) as follows.
\textcircled{\textbf{\small{1}}} EE tags the off-chip counter retrieval request on a miss to counter cache with a unique nonce. 
The nonce for a request is held in the MSHR entry for the counter cache miss request. 
\textcircled{\textbf{\small{2}}} The IVE uses the nonce in combination with a secret key (shared between the memory controller and the NVDIMM during authentication at boot) to encrypt the counter before sending it to the memory controller.
\textcircled{\textbf{\small{3}}} The memory controller decrypts the counter using the secret key and the nonce value 
\textcircled{\textbf{\small{4}}} After decrypting the counter, SMAC is computed and compared to the received one. The SMAC mismatch detects tampering, allowing {\TOOL} to identify the replay attack. 
Even if the attacker snoops on the memory bus and acquires the nonce value, it would fail to construct the encrypted counter value since it does not have the secret key.

\medskip
\noindent \textbf{Fine-grained integrity verification:}
By retaining all intermediate BMT nodes, {\TOOL} can precisely determine the tampered counter blocks covering 4KB of data block and intermediate nodes. If an incorrigible error occurs in Level h nodes but \texttt{Level\_h-1} nodes generate the identical root value (are verified), {\TOOL} can use \texttt{Level\_h-1} nodes to discover which \texttt{Level\_h} node is corrupted.

\medskip
\noindent \textbf{Tampering detection:}
BMT nodes fetched from the storage media are not trusted as they could have been tampered at storage or over the internal bus. 
First, the branch from the fetched node upto the BMT root or up to the first ancestor node cached in the BMT caches is verified.
Only then the BMT node is utilized to verify freshness of counter.
Thus, {\TOOL} can detect any tampering of the BMT nodes on storage media or over the internal NVDIMM bus. 

\subsection{\textbf{Data recovery}}
\medskip
\noindent \textbf{Power down or a crash:}
\label{subsec:power-down}
When the system detects a power failure, it sends a signal to the memory controller asserting the Asynchronous DRAM Refresh (ADR)~\cite{adr} mechanism as is done today.
The ADR would then transfer the WPQ entries from the memory controller to the NVDIMM using the stored energy. 
The PWRQ on the NVDIMM is used to queue writes from WPQ, and it also stores previously received writes that the IVE has not yet fully processed. On power down, 
the contents of PWRQ are written to a dedicated portion of the NVM media using a on-DIMM supercapacitor similar to ADR. 
Upon recovery, the pending requests in the PWRQ at the time of the power outage will be processed. 
In addition, the buffer designated for upper-level BMT nodes is written back to the NVM storage media. Following recovery, it is the responsibility of the IVE to validate unprocessed pending writes and to update BMT as described next.

\medskip
\noindent \textbf{Recovery process:}
\label{subsec:recovery}
{\TOOL}'s recovery time is short as it persists all security metadata. 
IVE must first guarantee that PWRQ entries (which were stored in a designated area of the NVDIMM during the power down) are untampered. IVE calculates the recursive hashes on these entries to generate the PHT and it's root. IVE securely retains the root of the PHT across  power cycles and uses it to verify against the computed PHT root on power up. Match between the two validates the integrity of the pending PWRQ requests following which, entries are loaded into the PWRQ and processed as detailed previously in \S\ref{subsec:IVE}. After these outstanding PWRQ entries are processed, the recovery operation is complete.

\subsection{\textbf{Hardware cost analysis}}
\label{hardware-cost-analysis}
IVE on the NVDIMM comprises $12$ hash units with $9$ employed for BMT updates and $3$ for PHT updates in a pipeline fashion.
{\TOOL} adds modest storage overheads.
PWRQ has $64$ entries, each having numerous fields ($5675$ bits), requiring a total of $44.34$KB of storage.
A PHT is constructed over the $64$ entries of PWRQ, resulting in $72$ nodes (each $64$B), needing an extra $4.5KB$. 
The upper-level BMT buffer contains the second and third highest level nodes and hence needs $4.5KB$. 
The lower-level BMT cache requires $32KB$.
Note that in the baseline, the BMT cache exist in the memory controller. 
The BTT consists of nine entries (each with $50$ bits) for pipelined inorder BMT updates.
BMT and PHT roots requires 128 bytes.
Therefore, each NVDIMM requires $85.52$KB of additional storage, which accounts for $\sim0.016 \%$ of total storage for a 512\,GB capacity NVDIMM.
On the memory controller, a nonce generation logic (random number generator) unit has been added. 
As in the baseline, there is a counter cache (32KB) and a MAC cache (32KB). In iMIV, the BMT cache has been relocated from the memory controller to the NVDIMM.

%% file: PaperContent/13_store_flow.tex
\subsection{\textbf{Putting it together: End-to-end flow for a write request}}
\label{subsec:store-request-analysis}

To summarise {\TOOL}'s architecture and operation, we depict the the end-to-end sequence of events for a write request that misses in both counter and SMAC cache as an example. The sequence of events are as follows (Figure~\ref{fig:Store-flow}).

\circled{\textbf{\small{1}}} A write request arrives at the memory controller.
\circled{\textbf{\small{2}}} EE looks up the counter and SMAC cache. 
\circled{\textbf{\small{3}}} On misses in both caches, nonce is generated to accompany the counter read request.
\circled{\textbf{\small{4}}} Counter and SMAC read requests are sent to the NVDIMM.
\circled{\textbf{\small{5}}} Counter is read from either internal NVDIMM buffers or media by the IVE.  \textcircled{\textbf{\small6}} Counter is verified using BMT.
\circled{\textbf{\small{7}}} IVE uses nonce and secret key in encrypting the counter.
\circled{\textbf{\small{8}}} Encrypted counter and SMAC are sent to the memory controller.
\circled{\textbf{\small{9}}} EE decrypts counter using nonce kept in MSHR and is incremented after SMAC verification.
\circled{\textbf{\small{10}}} OTP is generated using counter.
\circled{\textbf{\small{11}}} Plaintext is XORed with OTP to generate ciphertext.
\circled{\textbf{\small{12}}} New SMAC is computed using ciphertext and counter.
\circled{\textbf{\small{13}}} Ciphertext and SMAC write requests are sent to the NVDIMM.
\circled{\textbf{\small{14}}} On receiving the write request, IVE recalculates the SMAC after incrementing the counter retrieved locally from within the NVDIMM and compares it against that from the memory controller to ensure no tampering on the bus.
\circled{\textbf{\small{15}}} Afterward, IVE updates BMT reflecting the latest state of counters.
\circled{\textbf{\small{16}}} Once the security metadata and ciphertext are ready in PWRQ, they are drained to the NVM media via internal buffers.

%% file: PaperContent/8_methodology.tex
\begin{scriptsize}
\begin{table}[]
    \scriptsize
    \caption{Simulation configuration.}
    \vspace{-1em}
    \label{table:configuration}
    \begin{tabular}{|p{3cm}|p{4.8cm}|} 
    \hline
    \multicolumn{2}{|c|}{\textbf{Processor Cache Configuration}}\\
    \hline
    L1 Cache & 64KB, 4-way, 64B block \\ 
    L2 Cache & 512KB, 8-way, 64B block \\ 
    L3 Cache & 8MB, 64-way, 64B block \\ [1ex]
    \hline
    \multicolumn{2}{|c|}{\textbf{NVM (Intel Optane DC PMM) Parameters}}\\ [0.1ex]
        \hline
        Memory&512GB, 1333MHz clock\\
            &Media Read/Write Latency: 100/300ns\\
        RMW Buffer & 16KB\\
        AIT Buffer & 16MB\\
        \hline
    \multicolumn{2}{|c|}{\textbf{Memory Controller Parameters}}\\
    [0.1ex]
    \hline
    WPQ and RPQ & 12 entries (by default) \\
    \hline
    \multicolumn{2}{|c|}{\textbf{Encryption Engine Parameters}}\\
    [0.1ex]
    \hline
    Counter and SMAC cache & 32KB, 8-way, 64B block\\
    AES units and Hash units & 1 and 1\\
    AES and Hash Latencies & 80ns~\cite{aes_latency} and 40ns~\cite{aes_latency} (by default)\\
    \hline
    \multicolumn{2}{|c|}{\textbf{Integrity Verification Engine Parameters}}\\ [0.1ex]
    \hline
    BMT Lower Level Cache & 32KB, 8-way, 64B block \\ 
    BMT Upper Level Buffer & 4.5KB, 64B block\\
    BMT Hash units & 9, height of BMT (per-NVDIMM)\\
    PHT Hash units & 3, height of PHT \\
    Pending Write Request Queue & 64 entries\\
    Pending Read Request Queue & 16 entries\\
    BMT Tracking Table & 9 entries, height of BMT (per-NVDIMM)\\
    \hline
    \end{tabular}
\end{table}
\end{scriptsize}

\section{Evaluation}
\label{sec:evaluation} 

We extend the VANS simulator~\cite{vans} that mimics the internals of Intel's Optane NVDIMM and is verified against real Optane NVDIMMs.
We extend VANS to include an EE with a counter cache and a SMAC cache at the memory controller and the IVE with a split BMT cache at the NVDIMM. 
We capture memory access traces using Intel pintool~\cite{pintool} on a real Xeon-based system with Optane NVDIMMs and feed them into a simulator (integrated with CPU cache hierarchy) running in trace mode. Table~\ref{table:configuration} details the simulator parameter. 

We evaluate {\TOOL} against several prior works detailed in Table~\ref{table:evaluated_schemes}. For the Triad-NVM, we persist the lowest two levels of intermediate BMT nodes along with data, whereas for the ProMT\cite{promt}, we conservatively use the average effective height of $4.5$ for the BMT as analyzed in that work ($40$--$50$\% height reduction) based on hot and cold page detection.

\begin{scriptsize}
\begin{table}[]
  \centering
  \caption{Evaluated techniques with description.}
  \label{table:evaluated_schemes}
  \begin{tabular}{|p{1.4cm}|p{6.5cm}|}
        \hline
        \textbf{Technique} & \textbf{Description} \\
        \hline
        \texttt{\textbf{NS}} & No encryption and integrity verification  \\ 
        \hline
        \texttt{\textbf{BL}} & Baseline with encryption and integrity verification in MC \\
        \hline
        \texttt{\textbf{PLP}} & Persist-level parallelism, identical to baseline but utilizes hash computation pipeline for BMT node updates \\
        \hline
        \texttt{\textbf{Triad}} & Triad-NVM which is identical to baseline, but atomically persists only the lowest two levels of BMT in NVM  \\
        \hline
        \texttt{\textbf{ProMT}} & hot page detection and persisting of intermediate BMT nodes to either hotMT or globalMT \\
        \hline
        \texttt{\textbf{SBMF}} & Static BMF, reduces the effective height of BMT by two by caching level 3 nodes at memory-controller and does not persist intermediate BMT nodes\\
        \hline
        \texttt{\textbf{iMIV[-PLP]}} & Our approach with integrity verification in NVDIMMs that persists all intermediate BMT nodes, but do \emph{not} employ hash computation pipeline for BMT updates \\
        \hline
        \texttt{\textbf{iMIV}} &  Our approach with integrity verification in NVDIMMs that persists all intermediate BMT nodes and employs hash computation pipeline for BMT updates \\
        \hline
    \end{tabular}
\end{table}
\end{scriptsize}

\begin{scriptsize}
\begin{table}[]
    \centering
    \scriptsize
    \caption{Workloads and their NVM access characteristics.} 
    \label{table:workloads}  
    \begin{tabular}{|p{.8cm}|p{4.2cm}|C{.6cm}|C{1.2cm}|}        \hline
         & \textbf{Description} & \textbf{R:W ratio} & \textbf{Writes per 1K mem. access} \\
        \hline
        \texttt{\textbf{ARRSWP}} & Swapping random elements of a persistent array & 5.54 & 13.1\\ 
        \hline
        \texttt{\textbf{QUEUE}} & Enqueues and dequeues to a persistent queue & 10.28 & 1.89\\            
        \hline
        \texttt{\textbf{BST}} & Search, insertion and deletion to a persistent bst & 12.85 & 3.89\\            
        \hline
        \texttt{\textbf{BTREE}} & Insert and look up random elements in a persistent b-tree & 13.43 & 2.2\\
        \hline        
        \texttt{\textbf{CTREE}} & PMDK variant of crit-bit tree & 28.41 & 0.3\\
        \hline        
        \texttt{\textbf{HMAP}} & Hashmap implemented with PMDK & 6.28 & 2.2\\
        \hline
        \texttt{\textbf{REDIS}} & Persistent key-value store (PMDK) & 0.21 & 55.6\\
        \hline
        \texttt{\textbf{RBTREE}} & Insert and look up random elements in a persistent red-black tree & 13.45 & 1.39\\
        \hline
        \texttt{\textbf{RANDRW}} & Random updates to a persistent array & 3.07 & 9.3\\ 
        \hline        
        \texttt{\textbf{SEQRW}} & Sequential updates to persistent array & 2.07 & 13.9\\
        \hline        
    \end{tabular}
\end{table}
\end{scriptsize}

We evaluate workloads (Table~\ref{table:workloads}) that are implemented using PMDK library~\cite{pmdk}. 
Similar to prior works~\cite{janus,triad-nvm}, these include various persistent data structures such as array, queue and trees.
Workloads like \texttt{REDIS} and \texttt{ARRSWP} are write intensive with high NVM writes per 1K memory accesses and low read-write ratio.
For a comprehensive evaluation, we use \texttt{NVM-aware} and \texttt{NVM-agnostic} versions of the workloads.
In the former, the application allocates only selected data structures on the NVM as needed for recoverability, while  
\texttt{NVM-agnostic} places all the data on NVM.

%% file: PaperContent/9_evaluation.tex
\begin{figure*}[t!]
     \centering
     \begin{subfigure}[b]{\textwidth}
         \centering
         \includegraphics[width=0.4\linewidth]{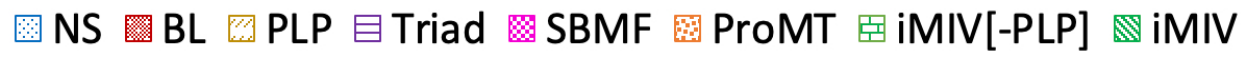}
     \end{subfigure}
     \hfill
     \begin{subfigure}[b]{\textwidth}
         \centering
        \includegraphics[width=.65\linewidth]{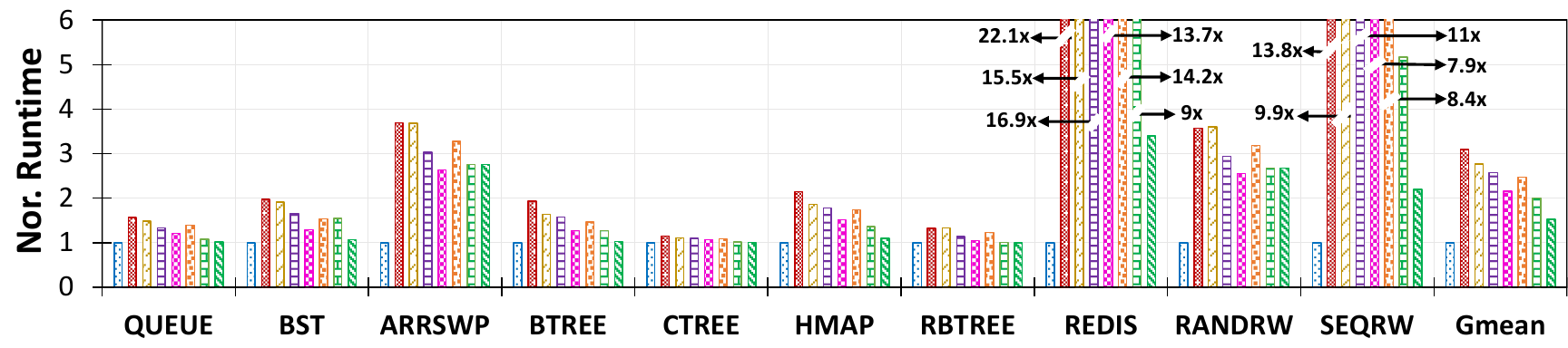}
         \caption{Analysis for \texttt{NVM-aware} version of workloads with strict persistency model employed.}
         \label{fig:Normalized_execution_SP_pm-aware}
     \end{subfigure}
     \hfill
     \begin{subfigure}[b]{\textwidth}
         \centering
         \includegraphics[width=.65\linewidth]{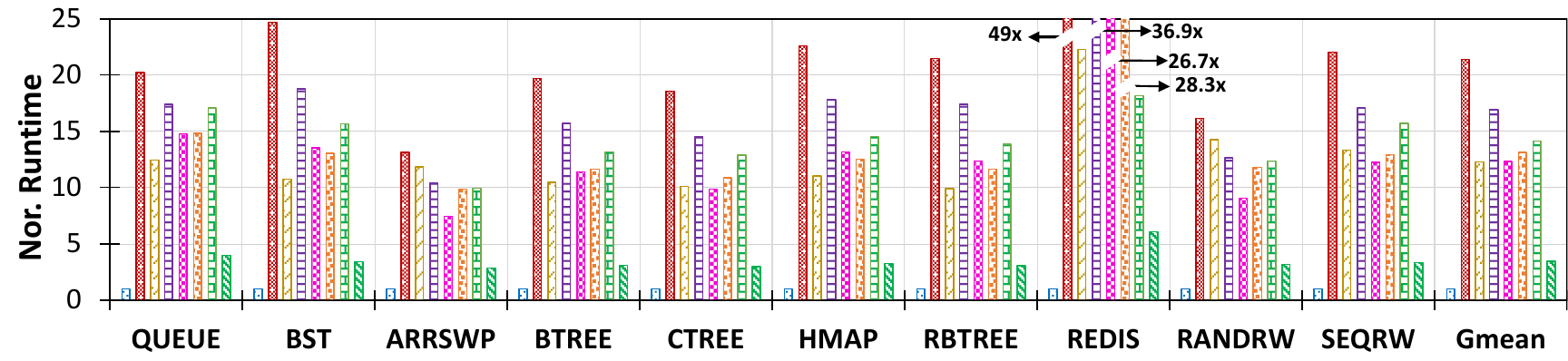}
         \caption{Analysis for \texttt{NVM-agnostic} version of workloads with strict persistency model employed.}
         \label{fig:Normalized_execution_SP_pm-agnostic}
     \end{subfigure}
     \hfill
     \begin{subfigure}[b]{\textwidth}
         \centering
         \includegraphics[width=.65\linewidth]{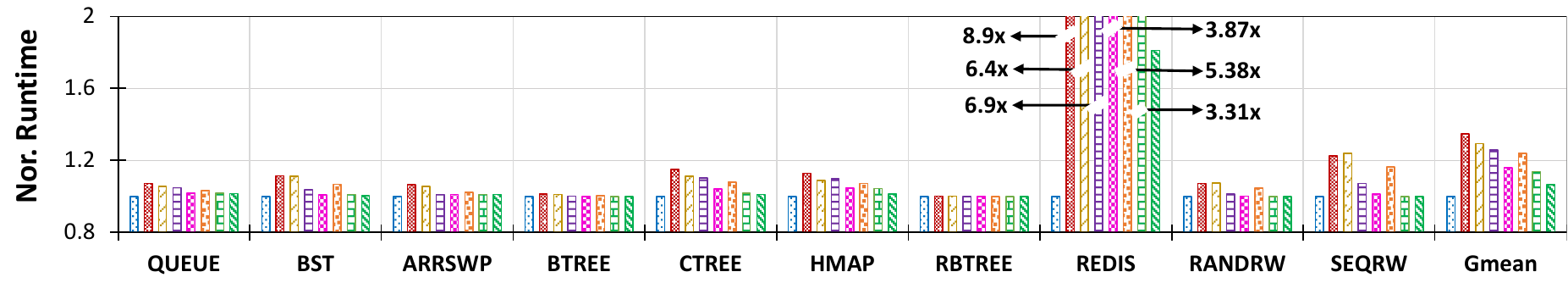}
         \caption{Analysis for \texttt{NVM-aware} version of workloads with x86 persistency model employed.}
         \label{fig:Normalized_execution_x86PM_pm-aware}
     \end{subfigure}
        \caption{Normalized runtime of various security techniques with no security (NS).}
        \label{fig:Normalized-execution}
\end{figure*}

\subsection{\textbf{Performance}}

\texttt{NVM-aware} versions are evaluated on two persistency models to demonstrate the wider applicability of our proposal. In
\emph{strict persistency model}~\cite{pmem_model} writes to NVM are persisted immediately. While \emph{x86 persistency model}~\cite{x86_pm_model}, is similar to today's Xeon systems with Optane NVDIMM.
Note that for \texttt{NVM-agnostic} versions, we use only the strict persistency model since they are agnostic to persistency, by definition.

\medskip
\noindent \textbf{Strict Persistency Model:}
\subsubsection{NVM-aware workloads} The strict persistency model for \texttt{NVM-aware} applications is accomplished by write-through CPU data and metadata caches.
WPQ ensures that data and security metadata are persisted together.
WPQ flushes the security tuple only when it is fully updated. 
On power failure, incomplete security tuples are discarded.

Figure~\ref{fig:Normalized_execution_SP_pm-aware} shows the normalized execution time with strict persistency model.
All numbers are normalized to \texttt{NS}.
Lower is better.
The \texttt{BL} persists the complete security tuple, including intermediate BMT nodes upon each NVM write.
Consequently, \texttt{BL} incurs a performance penalty of $205\%$, on average, over \texttt{NS}. 
Overheads are more for write-intensive workloads like \texttt{REDIS} (slows down by $22.1\times$) and \texttt{SEQRW} (slows down by $14.2\times$).
{\TOOL} significantly reduces overheads to $55\%$, on average. 
{\TOOL} reduced the overheads of \texttt{REDIS} from $22.1\times$ to $3.4\times$ and that of \texttt{SEQRW} from $14.2\times$ to $2.2\times$.

\texttt{PLP}~\cite{plp} performs better than \texttt{BL} ($12\%$ faster) by pipelining BMT updates. However, NVM bandwidth remains the bottleneck.
\texttt{Triad}~\cite{triad-nvm} persists only the two lowest levels of BMT to limit overheads and performs $20.3\%$ better than \texttt{BL}.
\texttt{SBMF} does not employ BMT update pipeline and does not persist intermediate BMT nodes, however, it reduces the height of BMT by caching the level 3 nodes at memory controller in a non-volatile cache.
{\TOOL} performs integrity verification in NVDIMM which alleviates bandwidth bottlenecks.
{\TOOL} performs $31\%$ faster than \texttt{{\TOOL}[-PLP]} demonstrating benefits of BMT pipelining.
Pipelining is more valuable when not limited by off-chip NVM bandwidth, as in {\TOOL}.
Overall, {\TOOL} achieves average speedup of $2\times$, $1.78\times$, $1.68\times$, $1.59\times$ and $1.38\times$ over \texttt{BL}, \texttt{PLP}, \texttt{Triad}, \texttt{ProMT} and \texttt{SBMF} respectively.

\medskip
\subsubsection{NVM-agnostic workloads} Figure~\ref{fig:Normalized_execution_SP_pm-agnostic} shows the performance impact for \texttt{NVM-agnostic} versions of the applications. 
The performance trends are similar to \texttt{NVM-aware} workloads but with a wider performance difference compared to \texttt{BL} as all data structures are placed on NVM. 
Notice that \texttt{ProMT} outperforms \texttt{{\TOOL}[-PLP]}, but not {\TOOL}.
\texttt{ProMT} reduces BMT update overheads while \texttt{{\TOOL}[-PLP]}, without the BMT hash pipeline, incurs additional overheads in updating BMT.
In \texttt{{\TOOL}[-PLP]}, BMT overheads delay processing and draining PWRQ entries which, in turn, delays the enqueuing of latter writes for \texttt{NVM-agnostic} workloads with large number of writes.
However, {\TOOL} with BMT hash pipeline performs $6.2\times$, $3.55\times$, $4.9\times$, $3.8\times$ and $3.57\times$ better than \texttt{BL}, \texttt{PLP}, \texttt{Triad}, \texttt{ProMT} and \texttt{SBMF}.

\medskip
\noindent \textbf{x86 Persistency Model:}
It employs writeback caches and applications use \texttt{clflush}/ \texttt{clwb} and \texttt{sfence} instructions to persist data in a crash-consistent manner\cite{x86_pm_model}.
The overheads (and thus, headroom for improvement) are less as the latency associated with NVM writes occurs only when the application flushes data to NVM and orders them with fence. However, {\TOOL} outperforms all techniques and outperfoms \texttt{BL} by an average of $26\%$ as shown in Figure~\ref{fig:Normalized_execution_x86PM_pm-aware}.

\begin{scriptsize}
\begin{figure*}[]
\centering
     \begin{subfigure}[b]{\textwidth}
         \centering
         \includegraphics[width=0.4\linewidth]{Graphs_PDF/Legend_Apr_19.pdf}
     \end{subfigure}
\includegraphics[width=.7\linewidth]{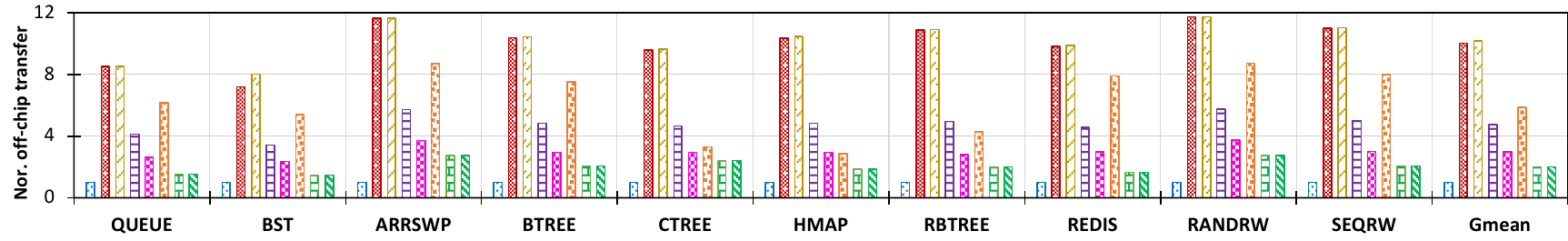}
\caption{\label{fig:SP_data_movement_analysis}Off-chip data transfer for \texttt{NVM-aware} workloads under strict persistency model.}
\end{figure*}
\end{scriptsize}

\subsection{\textbf{Off-chip data movement}}
The key advantage of {\TOOL} is its ability to avoid off-chip data movement due to integrity verification. 
Figure~\ref{fig:SP_data_movement_analysis} shows the off-chip data movement incurred by various techniques (normalized to \texttt{NS}) for strict persistency model.
\texttt{BL} and \texttt{PLP} incur the largest movement of data as they persist all intermediate BMT nodes. 
\texttt{ProMT} reduces data movement by leveraging hotMT to limit the number of persisted intermediate nodes.
\texttt{Triad} persists only the lowest two levels of the BMT whereas \texttt{SBMF} does not persist any intermediate nodes.
\texttt{SBMF} transfers counters off-chip, but {\TOOL} does not.
{\TOOL} reduces off-chip data movement by $5\times$ over \texttt{BL} and \texttt{PLP}, $2.34\times$, $2.9\times$ over \texttt{Triad} and \texttt{ProMT}, respectively with SPM.
Similar results are observed for x86 PM, with {\TOOL} lowering off-chip data movement by an average of $5.1\times$ over \texttt{BL} and \texttt{PLP}, $2.48 \times$ over \texttt{Triad}, $3.1\times$ over \texttt{ProMT} and $1.48\times$ over \texttt{SBMF}.

\subsection{\textbf{Recovery time}}
\begin{figure}[b!]
\centering
        \includegraphics[width=.8\linewidth]{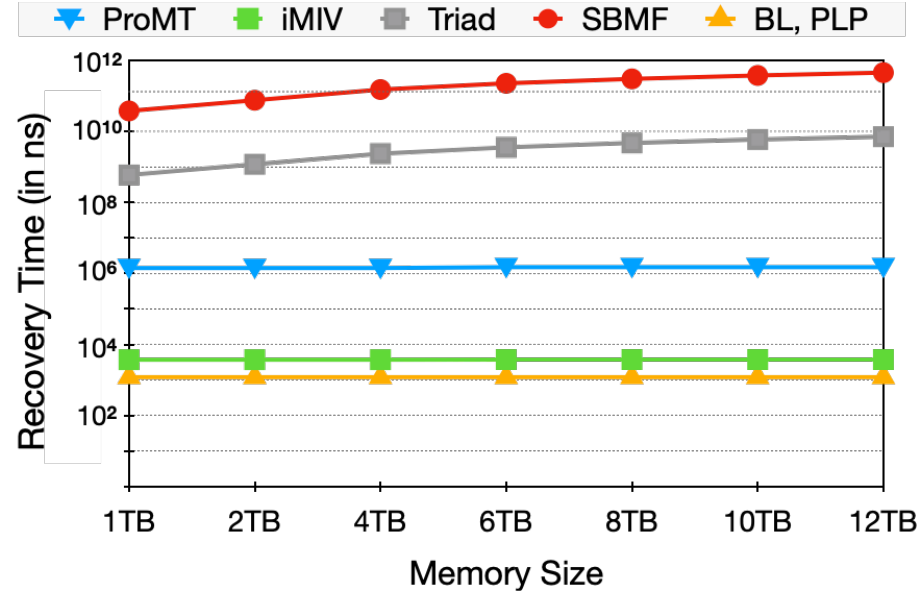}
      \caption{Recovery time analysis. y-axis is in log scale}
      \label{fig:recovery-time}
\end{figure}

Quickly verifying the security metadata is critical to recovery time.
Techniques that do not persist intermediate BMT nodes results in longer recover time 
mainly due to the reconstruction of BMT nodes~\cite{anubis}.
Figure~\ref{fig:recovery-time} shows the recovery time with increasing NVM capacity. To calculate the recovery time, we count the number of counter blocks and BMT nodes to be retrieved from the NVM media, and the time needed to reconstruct BMT (100ns for fetch and 40ns for hashing).
 
The baseline technique \texttt{BL} and \texttt{PLP} that persist all the intermediate nodes incurs little recovery time, but at the cost of high runtime overheads. 
Triad persists only the lowest two levels of BMT, necessitating reconstruction of the rest.
\texttt{ProMT} requires merging hotMT updates with the globalMT during recovery, and needs low recovery time.
Since \texttt{SBMF} does not persist intermediate nodes, and needs large recovery time.
Since {\TOOL} persists the intermediate nodes, BMT reconstruction is not needed. 
However, during recovery, the PWRQ entries which were computed but not written to the media at the time of the crash, should be written to the media. 
Also unprocessed PWRQ entries (at most, 64) needs processing by IVE on power up.
However, processing these entries adds little latency.
{\TOOL} performs on par with \texttt{BL} and \texttt{PLP} on recovery time with low overheads during execution.

\begin{figure}[b!]
\centering
        \includegraphics[width=.8\linewidth]{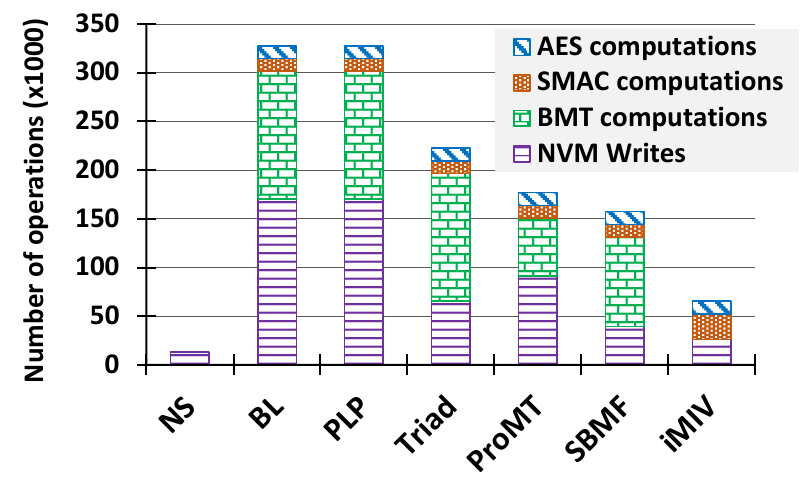}
    \caption{Number of operations performed for eADR system}
    \label{fig:eADR}
\end{figure}

\subsection{\textbf{Multi-DIMM scaling}}
We now examine the benefits of the per-NVDIMM IVE design of {\TOOL} in systems with multiple NVDIMMs.
We analyze the performance of different techniques with a single DIMM versus two DIMMs.
Across workloads, {\TOOL} beats prior works with average speedups of $1.3\times$, whilst the same for \texttt{ProMT, PLP, Triad}, and the baseline (\texttt{BL}) are $1.21\times$, $1.19\times$, $1.12\times$, and $1.1\times$ respectively.
As their accesses are consistently dispersed throughout the two DIMMs, \texttt{ARRSWP} and \texttt{RANDRW} provide significant gains with a $\sim1.85\times$ speedup for {\TOOL} while
the speedup for \texttt{ProMT, PLP, Triad}, and \texttt{BL} was $1.46\times$, $1.35\times$, $1.20\times$, and $1.24\times$ respectively.
This shows that per-NVDIMM IVE of \TOOL{} scales better with increasing DIMM counts.
\subsection{\textbf{Analysis for enhanced ADR (eADR) system}}
\label{subsec:eADR}

Intel's eADR~\cite{eadr} extends the persistence domain to include the entire CPU cache hierarchy, ensuring that dirty data is written back to NVM on power-down thus obviating the need to flush updates to NVM explicitly.
Intel concedes, however, that eADR is challenging to implement in practice~\cite{eadr-feasibility} as flushing all dirty cache blocks to NVM takes significantly more on-chip capacitor energy during power down.
Prior security techniques perform all security operations, including BMT updates, at memory controller and then persists security tuple, requiring enormous energy to flush the dirty cache lines at power down.

At power down, {\TOOL} encrypts dirty cache lines, computes SMAC at memory controller, and flushes them to NVM.
The flushing of cache lines can, however, overrun PWRQ (the default, $64$ entries).
We address this challenge by allocating a dedicated region within NVDIMM where the ciphertext and SMAC write requests that exceeded the PWRQ limit are stored. 
Further, these partial write requests need to be protected against tampering. 
The PHT protects only the PWRQ requests.
For thousands of requests, a PHT-style tree would consume too much energy.
We instead compute a \textit{single} hash value from all request contents (ciphertext and SMAC) and data block addresses and persist it in a secure register within the NVDIMM's IVE.
Later, on power up, BMT updates and SMAC verification are performed (\S\ref{subsec:power-down}) for the PWRQ entries.
Then, the hash verification is conducted on the overflow writes. 
Once verified, requests are loaded in batches into PWRQ and processed as normal write requests. 

Figure~\ref{fig:eADR} shows the number of operations that must be executed at a power down event for different techniques for an 8MB LLC in an eADR-enabled system, assuming $10\%$ of the cache lines were dirty (based on the behaviour of \texttt{NVM-aware} applications~\cite{whisper}).
{\TOOL} executes far fewer operations than alternatives (e.g., \texttt{BL} and \texttt{PLP} have $4.33\times$ more writes than {\TOOL}), hence decreasing the energy requirement.

\begin{figure}[]
\centering
\includegraphics[width=.8\linewidth]{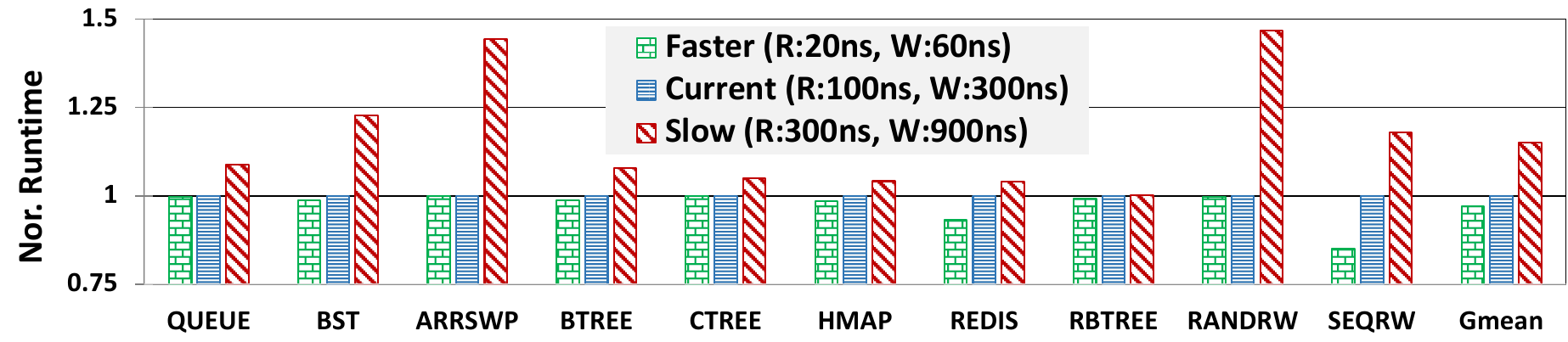}
\caption{\label{fig:NVM_media_latency_sensitivity}NVM media latency sensitivity analysis.}
\end{figure}

\subsection{\textbf{Sensitivity analysis}}

\noindent \emph{\textbf{Hash latency:}} 
Increase in hash latency has a linear impact on performance. 
Every $20ns$ increase in hash latency decreases performance by $\sim8\%$.

\medskip
\noindent \emph{\textbf{Read/write latency:}} {\TOOL}'s performance degrades by just $20\%$ on average even with $100ns$ to $300ns$ increase in read latency and $300ns$ to $900ns$ increase in write latency (Figure~\ref{fig:NVM_media_latency_sensitivity}).
This is due to the underlying design of Intel's Optane NVDIMMs, which employs layers of internal buffers~\cite{vans} (RMW and AIT) which coalesces writes to amortize the cost and hide higher media latency.

\medskip
\noindent \emph{\textbf{WPQ size:}}
Performance improves marginally (on avg. $4\%$) when the WPQ size is increased from 16 to 128. Near NVM integrity verification operation has alleviated pressure from WPQs as no intermediate BMT nodes gets enqueued.

\medskip
\noindent \emph{\textbf{Metadata cache size:}}
Smaller metadata cache size (from 128KB to 8KB) does not impact the performance significantly ($4\%$ on average). 
As {\TOOL} reduces the memory bandwidth consumed by the security metadata, one can afford to have few additional cache misses with smaller metadata cache.

%% file: PaperContent/10_Related_works.tex
\section{Related Works}
\label{sec:related_works}
Previous research efforts such as
iNVMM~\cite{iNVMM}, Covert~\cite{covert}, Acme~\cite{acme}, Secret~\cite{secret}, Deuce~\cite{deuce}, Morphable counters~\cite{morphable_counters},
SCA~\cite{selective_atomicity}, SecPM~\cite{secpm}, Triad-NVM~\cite{triad-nvm}, Janus~\cite{janus} developed secure NVM systems with varied security guarantees trading-off between performance, verifiability and recovery time.
SCA~\cite{selective_atomicity} integrated CME into an NVM system.
SecPM~\cite{secpm} proposed a write reduction technique to reduce the NVM writes.
SCA and SecPM did not focus on freshness of counters (by employing BMT).
Osiris~\cite{osiris} enhanced ECC to accommodate the counters and Anubis~\cite{anubis} emphasized on recovery time.
Janus~\cite{janus} decomposed security operations into a collection of concurrent sub-operations.
ORAM~\cite{oram} proposed secure DIMM with new protocols for side channel mitigation.
Triad-NVM~\cite{triad-nvm} emphasised the need of persisting intermediate BMT nodes to achieve finer verifiability guarantees by persisting lower levels of BMT.
PLP~\cite{plp} examined the ordering requirements of both data and the associated security metadata for proper crash recovery.
Shield-NVM~\cite{shield-nvm} presented an epoch-based technique with delayed spreading to decrease BMT hash calculations.
Recently, ProMT~\cite{promt} and BMF~\cite{bmf} reduced BMT overheads by using two BMTs rather than a single massive BMT and leveraged physical page access frequency.

%% file: PaperContent/11_conclusion.tex
\section{\textbf{conclusion}}
Fine-grain integrity verification, while desirable, is bottlenecked by limited off-chip bandwidth to NVDIMMs. We present {\TOOL}, an in-memory  fine-grained integrity verification technique 
that decreases off-chip data movement 
with low performance overheads and fast recovery which also seamlessly scales to systems with large aggregate NVM capacity.